\documentclass[12pt,a4paper]{article}
\usepackage{graphics}

\newcommand{\ovar}[1]{\overline{#1}}
\newcommand{\beq}{\begin{equation}}
\newcommand{\eeq}{\end{equation}}
\newcommand{\beqar}{\begin{eqnarray*}}
\newcommand{\eeqar}{\end{eqnarray*}}
\newcommand{\bra}[1]{\langle #1|}
\newcommand{\kket}[1]{| #1\rangle}

\begin{document}

\vspace*{.2cm}

\begin{center}
\begin{LARGE}{\bf %
$|V_{ub}|$ determination by $B \rightarrow D_s \pi$
}
\end{LARGE}\vskip4mm 

{\Large 
Hideaki Hayakawa\footnote{%
e-mail: haya@jodo.sci.toyama-u.ac.jp},  
Kaoru Hosokawa\footnote{%
e-mail: kaoru@jodo.sci.toyama-u.ac.jp}
and Takeshi Kurimoto \footnote{%
e-mail: krmt@sci.toyama-u.ac.jp}
}
\vskip3mm

Department of Physics, Faculty of Science,\\
Toyama University,\\
Toyama 930-8555, Japan\\

\vspace*{1cm}
\Large{\bf Abstract} 
\end{center}

We investigate $\ovar{B^0} \rightarrow D_s^- \pi^+$ decay  
in perturbative QCD approach which has recently been applied to 
$B$ meson decays. $\ovar{B^0} \rightarrow D_s^- \pi^+$ decay 
(and its charge conjugated mode)  
can be one of the hopeful modes to determine $|V_{ub}|$ since it
occurs through $b \rightarrow u$ transition only. We estimate 
both factorizable and non-factorizable contribution, and show 
that the non-factorizable contribution is much less than 
the factorizable one.
Our calculation gives 
$
\mbox{BR}(\ovar{B^0} \rightarrow D_s^- \pi^+) =
(50 \sim 70) \times f_{Ds}^2|{V_{ub}}{V_{cs}}|^2
$.  
\vskip7mm

\vskip2mm

 
\newpage

The determination of all the elements of Kobayashi-Maskawa 
matrix \cite{KM} is important for the consistency check of 
the standard model and a search for new physics. 
Much extensive experimental efforts have been being done 
at $B$ meson dedicated facilities (B factories) to 
complete the determination of KM matrix elements to third generation.
$|V_{cb}|$ has well been determined owing to heavy quark symmetry\cite{HQET}
to the accuracy of less than 10 \% error. But $|V_{ub}|$ has not yet been 
determined so precisely\cite{PDG}. The experimental error will be greatly 
reduced by the coming experiments in the very near future, while 
we need more effort to reduce the theoretical uncertainty.  
It is mainly due to hadronic effects which we do not yet have 
an precise method to calculate. 

So far $b \rightarrow u$ semi-leptonic 
decays have been mainly used for the experimental determination of $|V_{ub}|$. 
It is interesting to investigate other modes involving non-leptonic 
decays to extract $|V_{ub}|$ for the consistency checks of the 
experimental value of $|V_{ub}|$ and the theoretical methods 
to estimate hadronic 
effects. More experimental information can be available 
and we can tune up the theoretical methods. 
We propose here $\ovar{B^0} \rightarrow D_s^- \pi^+$ decay 
as one of good candidates to investigate.
The final state $D_s^- \pi^+$ is 
composed of $(s\bar c )$ $(u\bar d)$ quark state. The $b$ quark in 
$\ovar{B^0}$ meson cannot directly decay into $\bar c$ quark 
by $W^-$ emission as the color quantum number is different. 
Also no penguin, exchange nor annihilation contributions exist
since no $q \bar q$ state presents in the final state. Therefore, the 
decay occurs through $b \rightarrow u$ transition only, which makes 
$\ovar{B^0} \rightarrow D_s^- \pi^+$  
a good mode to determine $|V_{ub}|$. 

$\ovar{B^0} \rightarrow D_s^- \pi^+$ and its charge conjugate mode are 
hopeful from experimental point of view also.
Both $D_s$ and charged pion are relatively easy to be identified in 
the present experiments. Recently, BABAR and BELLE groups obtained 
the branching ratio,   
BR$(B^0 \rightarrow D_s^+ \pi^-) = 3.2 \pm 0.9 \pm 1.0 
 \times 10^{-5}$ (BABAR) \cite{BABAR},
$2.4 
{\scriptsize \begin{array}{l}
+1.0\\-0.8
\end{array}}
\pm 0.7 \times 10^{-5}$ (BELLE) \cite{KEKB}. 
With increasing statistics at B factories we can expect 
that the branching ratio is fixed more precisely
in the very near future.

The $\ovar{B^0} \rightarrow D_s^- \pi^+$ decay occurs through the effective 
Hamiltonian,
\begin{eqnarray}
{\cal H}_{eff} &=& \frac{G_F}{\sqrt{2}} 
{V_{ub}}{V_{cs}}^*
[c_1 (\bar s b)_{V-A}(\bar u c)_{V-A} + 
c_2 (\bar u b)_{V-A}(\bar s c)_{V-A}] + \mbox{(h.c.)} \\
&\equiv& \frac{G_F}{\sqrt{2}}
{V_{ub}}{V_{cs}}^*[c_1 O_1 +c_2 O_2] + \mbox{(h.c.)}, 
\end{eqnarray} 
where $c_{1,2}$ are the Wilson coefficients obtained by solving 
renormalization group equations,  
and $O_{1,2}$ are 4-quark operators\cite{BBL}.
We need to estimate $\bra{D_s^- \pi^+}O_{1,2}\kket{\ovar{B^0}}$ to obtain 
the branching ratio theoretically. So far, factorization ansatz\cite{BSW}
has often been used to estimate this kind of 2-body decay 
hadron matrix elements;
\begin{eqnarray}
A[\ovar{B^0} \rightarrow D_s^- \pi^+] 
&\simeq& 
\frac{G_F}{\sqrt{2}}
{V_{ub}}{V_{cs}}^*
\left(c_2 +\frac{c_1}{N_{eff}}\right)
\bra{D_s^-(P_2)}(\bar s c)_{V-A}\kket{0} 
\bra{\pi^+(P_3)}(\bar u b)_{V-A}\kket{\ovar{B^0}(P_1)} \nonumber\\
&=& i f_{Ds} \frac{G_F}{\sqrt{2}}
{V_{ub}}{V_{cs}}^*
\left(c_2 +\frac{c_1}{N_{eff}}\right)M_B^2
F_0({M_{Ds}}^2), \label{bswamp}
\end{eqnarray} 
where $N_{eff}$ is the effective color number 
and $f_{Ds}$ is the $D_s$ meson decay constant. 
Pion mass is neglected. 
The $B \rightarrow \pi$ transition form factors
are defined as follows;
\beq
\bra{\pi^+(P_3)}\bar u \gamma_\mu b\kket{\ovar{B^0}(P_1)}
\equiv \left[ (P_1 + P_3)_\mu - \left(\frac{M_B^2}{q^2}\right)q_\mu\right] 
      F_1(q^2) 
   + \left(\frac{M_B^2}{q^2}\right)
   q_\mu F_0(q^2), 
\eeq 
where $q = P_1 - P_3$.
Form factors
are to be obtained from another theory or experimental data.
The deviation of $N_{eff}$ from the number of color, $N_C = 3$,
accounts for so-called non-factorizable contributions.
With the amplitude given by eq.(\ref{bswamp}) 
the $\ovar{B^0} \rightarrow D_s^- \pi^+$ 
branching ratio is calculated as 
\beq
\mbox{BR}(\ovar{B^0} \rightarrow D_s^- \pi^+)
= \tau(\ovar{B^0})
\frac{(1-r^2)}{32\pi}
G_F^2 M_B^3 
\left(c_2 +\frac{c_1}{N_{eff}}\right)^2  F_0(M_{Ds})^2
f_{Ds}^2 |V_{ub} V_{cs}|^2,\label{brc}
\eeq
where $r\equiv M_{Ds}/M_B$.
Below we first give estimations of the branching ratio based 
on the factorization ansatz by using the $B \rightarrow \pi$ transition 
form factor from light-cone sum rules and lattice QCD. Then we
calculate ${\ovar{B^0}} \rightarrow D_s^- \pi^+$ amplitude 
by using perturbative QCD (pQCD) approach to estimate 
non-factorizable contribution.  
pQCD approach has been applied
to estimate pion electro-magnetic form factor, $B \rightarrow D$, 
$B \rightarrow \pi$ transition form factors and several decay amplitudes of 
2-body decays of B meson 
($D\pi$, $\pi\pi$, $K\pi$ and $K^*\gamma$) \cite{PQCD}.
The results of pQCD approach nicely agree with experimental data.
The advantage of pQCD approach lies in the point that the non-factorizable 
contribution can be calculated based on well-established perturbative 
QCD technique for heavy meson decays. 
We show that the non-factorizable contribution is about 10\% or less of 
the factorizable one in $B \rightarrow D_s\pi$, so that naive factorization 
estimation gives a reasonable prediction for this decay mode. 

In the evaluation of  eq.(\ref{brc}) we consider the Wilson coefficients 
at two scales, $\mu = M_B$ and $M_B/2$, to estimate the ambiguity coming 
from the choice of the scale;
\begin{center}
\begin{tabular}{c||c|c}
$\mu$  & $M_B$ & $M_B/2$ \\\hline
$c_1(\mu)$ & $-0.274$ & $-0.393$ \\
$c_2(\mu)$ & $1.12$ & $1.19$ 
\end{tabular} 
\end{center} 
The $B \rightarrow \pi$ transition form factor based on 
light-cone sum rules calculation can be parametrized as 
\beq
 F_0(q^2) = \frac{F_0(0)}{1 - a_F (q^2/M_B^2) +b_F (q^2/M_B^2)^2} 
\eeq
with $F_0(0) = 0.305$, $a_F = 0.266$ and $b_F=-0.752$~\cite{ball}.
With the calculation by lattice QCD\cite{UKQCD} we make a fit 
for $B \rightarrow \pi$ transition form factor as 
\beq
 F_0(q^2) = \frac{F_0(0)}{1 - c_F (q^2/M_B^2)}
\eeq
with $F_0(0) = 0.310$, $c_F = 0.760$.
The estimated value of 
$\mbox{BR}(\ovar{B^0} \rightarrow D_s^- \pi^+)/(f_{Ds}^2|V_{ub}V_{cs}|^2)$ is given 
in Table \ref{tablei}.
\begin{table}
\begin{center}
\begin{tabular}{c||cccc}
& 
$\begin{array}{c}
 N_{eff}=3\\ \mu=M_B
\end{array} $ 
&
$\begin{array}{c}
 N_{eff}=2\\ \mu=M_B
\end{array} $
&
$\begin{array}{c}
 N_{eff}=3\\ \mu=M_B/2
\end{array} $
&
$\begin{array}{c}
 N_{eff}=2\\ \mu=M_B/2
\end{array} $
\\\hline
lattice $F_0$ & 51.1 & 46.7 & 53.9 & 47.4 \\
sum rules $F_0$ & 44.0 & 40.1 & 46.4 & 40.8
\end{tabular} 
\end{center} 
\caption{$\mbox{BR}(\ovar{B^0} \rightarrow D_s^- \pi^+)/(f_{Ds}^2|V_{ub}V_{cs}|^2)$ }
\label{tablei}
\end{table} 
By adopting $f_{Ds} = 0.241 \pm 0.032$ GeV obtained by taking average of
several experimental data\cite{PRS}, 
we can summarize the naive factorization estimation as 
\beq
\mbox{BR}(\ovar{B^0} \rightarrow D_s^- \pi^+)/|V_{ub}V_{cs}|^2  
=
\left\{
\begin{array}{cl}
2.0 \sim 4.0 & \mbox{(lattice $F_0$)}\\
1.8 \sim 3.5 & \mbox{(sum rules $F_0$)}
\end{array} 
\right. .
\eeq
The ambiguity from the choice of 
$N_{eff}$ and $\mu$ 
is about 10\% and 5\%, respectively.
The major ambiguity lies in $f_{Ds}$ value.

\begin{figure}[h]
\centerline{
\resizebox{14cm}{!}{
\includegraphics{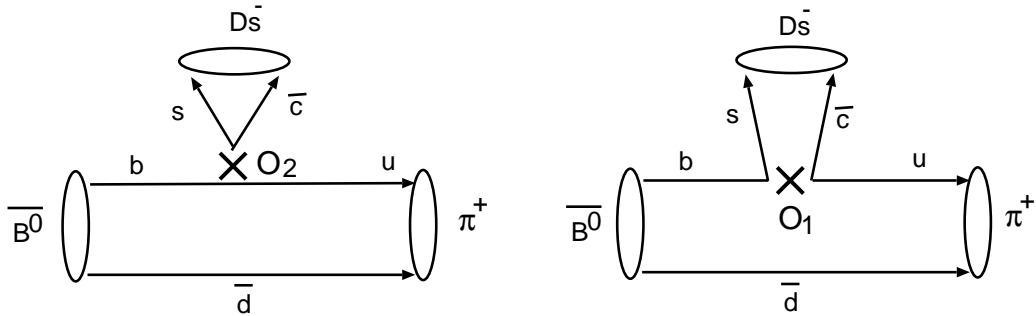}
 }
}
\caption{Diagrams contributing to $\ovar{B^0} \rightarrow D_s^- \pi^+$. 
Gluon is not shown.}
\label{dspifig}
\end{figure}

There are factorizable and non-factorizable contribution in 
$\ovar{B^0} \rightarrow D_s^- \pi^+$. When a gluon connects the spectator 
$\bar d$ quark and a quark in $D_s$ meson 
in Fig.\ref{dspifig}, the contribution cannot be 
factorized as in eq.(\ref{bswamp}).
The non-factorizable contribution is taken into account by changing 
the color number from $N_C = 3$ to $N_{eff}$ in the calculation based on 
the factorization ansatz. 
However, the number to be taken as $N_{eff}$ is theoretically unclear.  
We have to rely on fit of $N_{eff}$ by using experimental data other 
than $B\rightarrow D_s\pi$. But, we cannot simply adopt 
the $N_{eff}$ in other decays because the topology of diagrams contributing to 
other decays is not necessary same as in the case of 
$\ovar{B^0} \rightarrow D_s^- \pi^+$ decay. 
Here we calculate $\ovar{B^0} \rightarrow D_s^- \pi^+$ amplitude in 
pQCD approach\cite{PQCD} to estimate the non-factorizable contribution 
based on QCD.

\begin{figure}
\centerline{
\resizebox{13cm}{!}{
\includegraphics{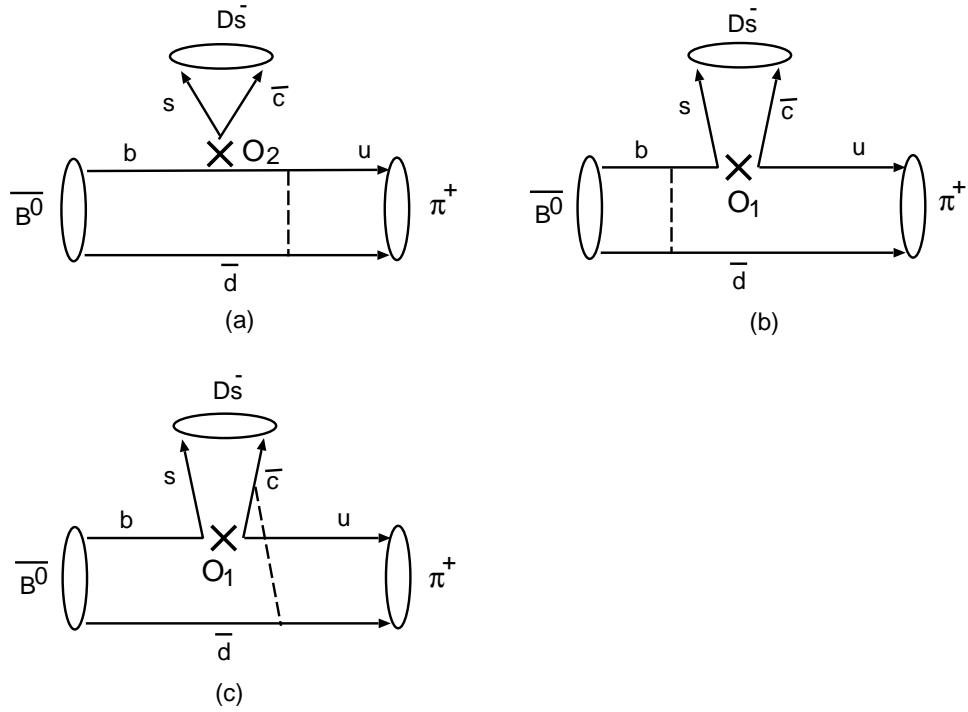}
 }
}
\caption{Some 
diagrams contributing to $\ovar{B^0} \rightarrow D_s^- \pi^+$ in pQCD.
Gluon is shown in dashed line.}
\label{pqcdfig}
\end{figure}
\begin{figure}
\centerline{
\resizebox{8cm}{!}{
\includegraphics{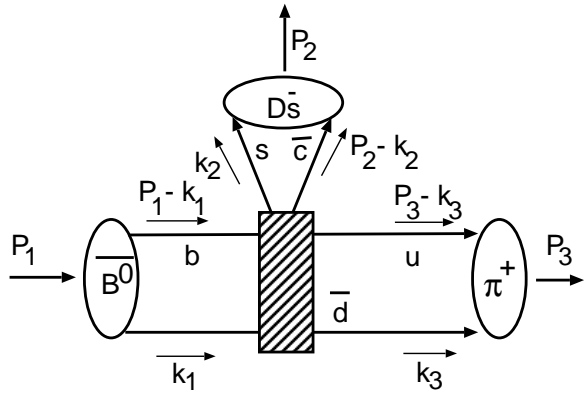}
 }
}
\caption{ Momentum flow in the  
diagrams contributing to 
$\ovar{B^0} \rightarrow D_s^- \pi^+$ in pQCD.
}
\label{mom}
\end{figure}
Some of representative diagrams contributing to 
$\ovar{B^0} \rightarrow D_s^- \pi^+$ 
in pQCD approach is shown in Fig.\ref{pqcdfig}. The point is that 
in 2 body decays of $B$ meson the spectator quark has to obtain high  
momentum to form a meson with one of the emitted rapid quarks 
from $b$ quark decay. 
The high momentum is carried by a gluon, so the perturbative 
QCD treatment is possible.
The non-perturbative nature is put into meson wave functions.
For more details refer the papers in \cite{PQCD}. 
There is another approach of calculating non-leptonic 2-body decays of 
B meson by using perturbative QCD, the so-called QCD factorization 
method\cite{BBNS}. One of the major differences between pQCD approach and
QCD factorization lies in the treatment of the transverse momentum. 
It is taken into account in the pQCD approach while neglected in the 
QCD factorization. Without the transverse momentum the diagrams 
given in fig.\ref{pqcdfig} 
have infrared singularities which makes the results of the 
calculation implausible.
The calculation based on pQCD approach can be done 
in a similar way to calculate $B \rightarrow D\pi$ amplitude 
given in Li and Meli\'c\cite{PQCD}. 
The assignment of momenta is shown in Fig.\ref{mom} where 
$P_1 = (M_B/\sqrt{2})\: (1,1,0_T)$, 
$P_2 = (M_B/\sqrt{2})\: (r^2,1,0_T)$,
$P_3 = (M_B/\sqrt{2})\: (1-r^2,1,0_T)$ and
$k_1 = (0,x_1(M_B/\sqrt{2}),k_{1T})$,
$k_2 = (0,x_2(M_B/\sqrt{2}),k_{2T})$,
$k_3 = (x_3 (1-r^2)(M_B/\sqrt{2}),0, k_{3T})$ 
in light-cone coordinate.
We obtain
\beq
\Gamma[\ovar{B^0} \rightarrow D_s^- \pi^+] =
\frac{G_F^2}{128\pi}|V_{ub}V_{cs}|^2 M_B^3 \frac{(1-r^2)^3}{r}
|f_{Ds}\xi_{\rm int} + {\cal M}_{\rm int}|^2.   
\eeq
The factorizable contribution, $f_{Ds}\xi_{\rm int}$, is given as
\begin{eqnarray}
\xi_{\rm int}&=&16\pi C_F\sqrt{r}M_B^2
\int_0^1 dx_1dx_3\int_0^{1/\Lambda}b_1db_1b_3db_3
\phi_B(x_1,b_1) 
\nonumber \\
& &\times \alpha_s(t_{\rm int})\left(c_2(t_{\rm int}) + \frac{c_1(t_{\rm int})}{N_C}\right) 
\exp[-S_B(t_{\rm int})-S_\pi(t_{\rm int})]
\nonumber \\
& &\times \Biggl[ [
\{1+x_3(1-r^2)\} \phi_A(x_3) 
+ r_0 \{ \left(\frac{1+r^2}{1-r^2}\right)
      - 2x_3 \}\phi_P(x_3) \nonumber \\
&& \ \ \ \ \  
- r_0 ( 2x_3 -1) \phi_T(x_3) 
]
h(x_1,x_3,b_1,b_3,m_{\rm int})
\nonumber \\
& &\ \ + [x_1r^2 \phi_A(x_3)
  + 2r_0 \{ 1 - \left( \frac{r^2}{1-r^2} \right)x_1 \}\phi_P(x_3) ]
h(x_3,x_1,b_3,b_1,m_{\rm int})\Biggr]\;,
\label{int} 
\end{eqnarray}
where $\exp[-S_B(t_{\rm int})-S_\pi(t_{\rm int})]$ is Sudakov factor\cite{PQCD},
$m_{\rm int}=(1-r^2)M_B^2$, $\Lambda = \Lambda_{\rm QCD}$ and 
$t_{\rm int}={\rm max}(\sqrt{x_1m_{\rm int}},\sqrt{x_3m_{\rm int}},
1/b_1,1/b_3)$. The parameter $b_i$ ($i=1\sim 3$) is the conjugate 
variable of $k_{iT}$.  
The function $h(x_1,x_3,b_1,b_3,m)$ is defined as
\begin{eqnarray}
h(x_1,x_3,b_1,b_3,m)&=&S_t(x_3) K_{0}\left(\sqrt{x_1x_3m}b_1\right)
\nonumber \\
& &\times \left[\theta(b_1-b_3)K_0\left(\sqrt{x_3m}
b_1\right)I_0\left(\sqrt{x_3m}b_3\right)\right.
\nonumber \\
& &\ \ \ \left.+\theta(b_3-b_1)K_0\left(\sqrt{x_3m}b_3\right)
I_0\left(\sqrt{x_3m}b_1\right)\right],
\end{eqnarray}
where 
\beq
  S_t(x) = \frac{2^{1+2c} \Gamma(c +3/2)}{\sqrt{\pi}\Gamma(c+1)} [x(1-x)]^c 
\ \ (c = 0.3 \sim 0.4), 
\eeq
which comes from threshold resummation\cite{KLS}.
The non-factorizable contribution is given as
\begin{eqnarray}
{\cal M}_{\rm int}&=& -32\pi\sqrt{2N_C} C_F\sqrt{r}M_B^2
\int_0^1 dx_1 dx_2 dx_3\int_0^{1/\Lambda}b_1 db_1 b_2 db_2
\phi_B(x_1,b_1)\phi_{Ds}(x_2,b_2)
\nonumber \\
& &\times \alpha_s(t_d)\frac{c_1(t_d)}{N_C}\exp[-S(t_d)|_{b_3=b_1}] 
\nonumber \\
& &\times 
\Biggl[[
  (x_1-x_2)(1+r^2)\phi_A(x_3) 
 +r_0\{x_3 -\frac{(x_1 - x_2)r^2}{(1-r^2)} \}\phi_P(x_3) 
\nonumber \\
&& \ \ \ \ \ 
 + r_0\{x_3 + \frac{(x_1 - x_2)r^2}{(1-r^2)} \}\phi_T(x_3)
]S^{(1)}_{\rm int} h^{(1)}_d(x_i,b_i)
\nonumber \\
& &\hspace{0.5cm}
+[\{ 1 -x_1 -x_2 +(1-r^2)x_3)\}\phi_A(x_3) 
  -r_0 \{x_3 -\frac{(2+x_1+x_2)r^2}{(1-r^2)}  \}\phi_P(x_3) 
\nonumber \\
&& \ \ \ \ \ \  \
  -r_0 \{x_3 +\frac{(x_1+x_2)r^2}{(1-r^2)} \}\phi_T(x_3) 
]S^{(2)}_{\rm int} h^{(2)}_d(x_i,b_i)
\Biggr]\; ,
\end{eqnarray} 
where $N_C =3$, $t_d={\rm max}(DM_B,\sqrt{|D_1^2|}M_B,\sqrt{|D_2^2|}M_B,
1/b_1,1/b_2)$,
\begin{eqnarray}
D^{2}&=&x_{1}x_{3}(1-r^{2})\;,
\nonumber \\
D_{1}^{2}&=&(x_{1}-x_{2})x_{3}(1-r^{2})\;,
\nonumber \\
D_{2}^{2}&=&(x_{1}+x_{2})r^{2}-(1-x_{1}-x_{2})x_{3}(1-r^{2})\;.
\end{eqnarray}
The functions $h_d^{(1)}$ and $h_d^{(2)}$ are defined as 
\begin{eqnarray}
h^{(j)}_d&=& \left[\theta(b_1-b_2)K_0\left(DM_B
b_1\right)I_0\left(DM_Bb_2\right)\right. \nonumber \\
& &\quad \left.
+\theta(b_2-b_1)K_0\left(DM_B b_2\right)
I_0\left(DM_B b_1\right)\right]\;  \nonumber \\
&  & \times \left( \begin{array}{cc}
 K_{0}(D_{j}M_Bb_{2}) &  \mbox{for $D^2_{j} \geq 0$}  \\
 \frac{i\pi}{2} H_{0}^{(1)}(\sqrt{|D_{j}^2|}M_Bb_{2})  &
 \mbox{for $D^2_{j} \leq 0$}
  \end{array} \right)\;.          
\end{eqnarray} 
The factor $S_{\rm int}$ accounts for threshold resummation in 
non-factorizable contribution investigated in the work by Li and Ukai\cite{liukai}. 
The effects of this factor shall be discussed later.   

We have included twist 3 component into wave functions;
\begin{eqnarray}
 B(P) &=& [\not P + M_B ]\gamma_5 \phi_B(x),\\
 {D_s}(P)  &=& \gamma_5[\not P + M_{Ds} ]\phi_{Ds}(x),\\
 \pi^+(P) &=&
\gamma_5 [\not P \phi_A(x) + m_0 \phi_P(x)
     - m_0(\not v\not n -v\cdot n)\phi_T(x)],
\end{eqnarray}
where $m_0 \equiv M_\pi^2/(m_u + m_d)$, $r_0 \equiv m_0/M_B$,
$v = (1,0,0_T)$ and $n=(0,1,0_T)$ in light-cone coordinate.
We have adopted the following functions
for $B$ meson and pion;

\begin{eqnarray}
\phi_B(x,b) &=& N_B x^2 (1-x)^2 
\exp\left[-\frac{1}{2}\left(\frac{xM_B}{\omega_B}\right)^2 
 -\frac{1}{2} (\omega_B'b)^2 
\right], \\
\phi_A(x) &=& \frac{3f_\pi}{\sqrt{2N_C}}x(1-x)[1 + a_{2} C_{2}^{3/2}(1-2x)
             + a_{4} C_{4}^{3/2}(1-2x)],\\
\phi_P(x) &=& \frac{f_\pi}{2\sqrt{2N_C}}[1 + a_{2p} C_{2}^{1/2}(1-2x) 
           + a_{4p} C_{4}^{1/2}(1-2x)],\\
\phi_T(x) &=& \frac{f_\pi}{2\sqrt{2N_C}}(1-2x)[1 + 
 6a_{2t}(10x^2 -10x +1)],
\end{eqnarray}
where $N_B$ is normalization constant to have 
$\int_0^1 \phi_B(x,0)dx = f_B/2\sqrt{6}$, and $C_j^k$ is a 
Gegenbauer polynomial: 
$C_2^{1/2} (x)=(1/2)(3x^2 -1)$, 
$C_4^{1/2} (x)=(1/8)(35x^4 -30 x^2 +3)$,
$C_1^{3/2} (x)=3x$,
$C_2^{3/2} (x)=(3/2)(5x^2 -1)$,
$C_4^{3/2} (x)=(15/8)(21x^4 -14 x^2 +1)$.
The parameters in the pion wave functions are
given in ref.\cite{Ball};
\beq
a_2 = 0.44, \ \ 
a_4 =  0.25, \ \ 
a_{2p} = 30 \eta_3, \ \ 
a_{4p} = -30\eta_3\omega_3 , \  \
a_{2t} = 5\eta_3 -\frac{1}{2}\eta_3\omega_3, 
\eeq
with $\eta_3 = 0.015$ and $\omega_3 = -3.0$. 
As for $D_s$ meson wave function we take;
\beq
\phi_{Ds}= N_D x(1-x)[ 1 + \frac{c_{d}}{3}C_1^{3/2}(1-2x)].
\eeq

By analyzing $B \rightarrow \pi$ and 
$B \rightarrow D$ form factors  
we have obtained a set of parameters consistent with experimental 
data and other theoretical predictions\cite{KLS,KHS};
$$
\omega_B =\omega_B' = 0.4  \mbox{ for $B$ meson and }
c_{d} =0.7  \mbox{ for $D$ meson}.
$$
The $D_s$ meson wave function is thought to be similar to $D$ meson wave function 
by $SU(3)$ flavor symmetry. The parameter $c_{d}$ are 
varied in the numerical analysis to see their effects;  
$$
c_{d} =   0.5,\  0.7,\  0.9.
$$
The numerical results on the $\ovar{B^0} \rightarrow D_s^- \pi^+$ branching ratio 
is given in Table \ref{brtbli} by taking $c=0.35$ and $S_{\rm int}^{(1,2)} =1$,
i.e. without the threshold resummation effect 
in non-factorizable contribution. 
The results show that the branching ratio of this mode is almost
insensitive to the $D_s$ meson wave function for a reasonable range of
parameter $c_d\;$\cite{KHS}. 
\begin{table}
\begin{center}
\begin{tabular}{c||ccc}
$c_d$    & 0.5 & 0.7 & 0.9  \\\hline
$|{\cal M}_{\rm int}|/ f_{Ds}\xi_{\rm int} $ &0.093 & 0.088 & 0.082\\
${\rm BR}/{(f_{Ds}^2|V_{ub}V_{cs}|^2)} $ & 60 & 59 & 59 
\end{tabular}
\end{center}
\caption{$\mbox{BR}(\ovar{B^0} \rightarrow D_s^- \pi^+)/|V_{ub}V_{cs}|^2$ and the
ratio of non-factorizable contribution to factorizable one
for different $D_s$ meson wave function parameter $c_d$.
$f_{Ds}$ should be given in GeV.}
\label{brtbli}
\end{table}

The effect of the threshold resummation is also investigated. 
The threshold resummation parameter $c$ in eq.(12) for the 
factorizable contributions is varied within the range consistent 
with the $B\rightarrow\pi$ form factor analysis\cite{KLS}. 
As for the non-factorizable contribution we take 
$S_{\rm int}^{(1)}= S_t(x_3)$ and $S_{\rm int}^{(2)}= 1$
following the arguments given in ref.\cite{liukai}.
The parameter $c$ is taken to be the same with that for 
factorizable contributions for the simplicity of the calculation. 
Two kinds of calculations have been made with or without 
threshold resummation factor in the non-factorizable contribution. 
The results are shown in Table \ref{brtblii}.
It is found that the branching ratio varies about 15\% depending on the 
choice of the parameter $c$. But this parameter is just a numerical convenience 
to fit the true form of the following threshold resummation factor 
by eq.(12);
\beq
S_t(x) = \int_{a-i\infty}^{a-i\infty} \frac{dN}{2\pi i} \frac{J(N)}{N} (1-x)^{-N}, 
\eeq
where $a$ is an arbitrary real constant larger than all the real parts of the poles 
in the integrand and
\beq
J(N) = \exp\left[
\frac{1}{2}\int_0^1 dz \frac{1-z^{N-1}}{1-z} 
\int_{(1-z)}^{(1-z)^2}\frac{d\lambda}{\lambda} \gamma_K(\alpha_s) 
\right]
\eeq
with $\gamma_K(\alpha_s)$ being the anomalous dimensions\cite{liukai,KHS}.
Thus the parameter $c$ is absent in the more careful (but complicated) 
treatment of numerical calculation where the above equations are used. 
So this ambiguity does not lead to a true theoretical error.
The effect of the existence of the threshold resummation in 
non-factorizable contribution is found to be small in this decay. 
This is because non-factorizable contribution is small, 10\% or less 
in amplitude, in comparison with the factorizable one in this 
decay mode. But the effects of the threshold resummation in 
non-factorizable part can be significant in another modes such as 
color-suppressed decay like $B^0 \rightarrow \ovar{D^0}\pi^0$\cite{KLL}. 
\begin{table}
\begin{center}
\begin{tabular}{cc||ccc}
&$c$    & 0.30 & 0.35 & 0.40  \\\hline
(a)& ${\cal M}_{\rm int}/ f_{Ds}\xi_{\rm int} $ & 
$ 0.079 - 0.022i$&$0.085 - 0.024 i$&$0.090 - 0.026 i$ \\
&${\rm BR}/{(f_{Ds}^2|V_{ub}V_{cs}|^2)} $ & 68 & 59 & 52\\\hline
(b)& ${\cal M}_{\rm int}/ f_{Ds}\xi_{\rm int} $ & 
$0.069 + 0.018i$&$0.074 + 0.025 i$&$0.077 + 0.031 i$ \\ 
&${\rm BR}/{(f_{Ds}^2|V_{ub}V_{cs}|^2)} $ & 66 & 58 & 51 
\end{tabular}
\end{center}
\caption{
(a) $\mbox{BR}(\ovar{B^0} \rightarrow D_s^- \pi^+)/|V_{ub}V_{cs}|^2$ 
and the ratio of non-factorizable contribution to factorizable one 
for different threshold resummation parameter $c$ without threshold resummation 
in non-factorizable contribution. (b) same as (a) with threshold resummation
in non-factorizable contribution.}
\label{brtblii}
\end{table}

The prediction by pQCD approach is given  considering the ambiguity 
discussed before as 
\begin{eqnarray}
\mbox{BR}(\ovar{B^0} \rightarrow D_s^- \pi^+) &=& 
(50 \sim 70 )  \times f_{Ds}^2|{V_{ub}}{V_{cs}}|^2\\
&=& (2.4\sim 4.6) \times |{V_{ub}}{V_{cs}}|^2  .
\end{eqnarray} 
If we take the central values of parameters and the experimental data 
of $B\rightarrow D_s \pi$ branching ratio given by BABAR and BELLE, we obtain,
\beq
|V_{ub}| = ( 3\pm 1 ) \times 10^{-3},
\eeq 
which is in good agreement with the value of $|V_{ub}|$ obtained in 
$b\rightarrow u$ semi-leptonic decay\cite{PRD}.
This agreement gives a support to our treatment of $B$ meson decays 
in pQCD approach. It also implies that the naive factorization ansatz 
works well also in this decay mode\cite{CLJ} since our calculation has shown 
the dominance of factorizable contribution.
The ambiguity will be reduced when $f_{Ds}$ value is fixed more precisely in 
the future experiments.
In the near future high statistics of $B$ meson decay data will be available, 
then we can make pQCD prediction more precise by fitting parameters of the 
wave functions with rich experimental data. Then $|V_{ub}|$ can be determined as 
precise as those from semi-leptonic decay by using 
the branching ratio of $\ovar{B^0} \rightarrow D_s^- \pi^+$.

\vskip10mm

\centerline{\large\bf Acknowledgements}
The authors thank Prof. M.~Yamauchi at KEK for pointing out the 
possibility of $|V_{ub}|$ determination by $\ovar{B^0} \rightarrow D_s^- \pi^+$.
T.K thanks the members of pQCD working group for fruitful discussions and
encouragement. 
The work of T.K was supported in part by Grant-in Aid for Scientific
Research from the Japan Society for the Promotion of Science
under the Grant No. 11640265.

\end{document}